\documentstyle[prd,aps,epsf,epsfig]{revtex} 
\begin{document}
\draft
\twocolumn[\hsize\textwidth\columnwidth\hsize\csname@twocolumnfalse\endcsname

\title{Pattern formation and selection in quasi-static fracture}

\author
{{Kwan-tai Leung}\cite{email_ktl}${}^1$ 
and {Zoltan N\'eda}\cite{email_zn}${}^2$}
\address{
${}^1$ Institute of Physics, Academia Sinica,
Taipei, Taiwan 11529, R.O.C.\\
${}^2$ Department of Theoretical Physics, Babe\c{s}-Bolyai University,\\
str. Kogalniceanu  nr.~1, RO-3400 Cluj-Napoca, Romania
}

\maketitle
\centerline{\small (Last revised \today)}

\begin{abstract}
Fracture in quasi-statically driven systems is studied by means of 
a discrete spring-block model. 
Developed from close comparison with desiccation experiments,
it describes crack formation induced by friction on a substrate.
The model produces cellular, hierarchical patterns of cracks,
characterized by a mean fragment size linear in the layer thickness,
in agreement with experiments.
The selection of a stationary fragment size is explained by exploiting
the correlations prior to cracking.
A scaling behavior associated with the thickness and substrate coupling,
derived and confirmed by simulations, 
suggests why patterns have similar morphology
despite their disparity in scales.
\end{abstract}

\vspace{2pc}
]



Nature is full of fascinating patterns\cite{stevens}.
A ubiquitous yet relatively less
explored class of patterns is that produced
by the fracture of solids\cite{lawn}, as often seen 
in cracked structures, battered roads and dried out fields.
Although there have been early observations
and characterizations\cite{stevens,playa,Walker},
only rather recently has it been studied systematically 
\cite{recent,groisman,BBN,la97,kitsunezaki}.
The mathematical problem of a network of interacting 
cracks\cite{Kachanov} is a formidable one, in view of
the difficulties facing
a lone crack propagating in a homogeneous medium\cite{CZM}.
One way to make progress is to turn to a statistical description.
This view has been pursued extensively in investigations of 
the fracture of disordered media\cite{HR},
borrowing concepts such as percolation and universality
from studies of phase transitions.
For crack patterns,
an immediate observation is the geometrical
similarities of patterns over a wide range of scales,
from microns\cite{recent} to kilometers\cite{playa}.
This suggests some universal mechanism is at work
and microscopic details may be unimportant.
Hence, analogous to phase transitions,
a mesoscopic, coarse-grained description may be sufficient
and more useful than a microscopic one, 
provided essential features are captured.

Crack patterns often arise
from slow physical (e.g., embrittlement, contraction) 
or chemical (e.g., oxidation) variations,
or their combination, in the material properties of 
an overlayer on top of an inert substrate.
The overlayer may fail in different ways\cite{Hutchinson}, such as 
decohesion, buckling, spalling and in-plane cracking.
In this paper, we 
address the pattern-selection aspects of
quasi-static, in-plane cracking  by 
identifying the dominant mechanism and control parameters
that determine 
the ensuing patterns. 
By understanding a simplified, coarse-grained model, we hope to achieve
the same for the phenomena in general.


{\em A Bundle-Block Model\/} --
In a mesoscopic approach, 
the grains in the overlayer are represented by an array of blocks.
Each pair of {\em neighboring\/} blocks is connected by a bundle of 
$H$ bonds (coil springs) each of which has
spring constant $k\equiv 1$ and relaxed length $l$.
Initially, the blocks are randomly displaced by 
$\vec{r}=(x,y)$, where $|\vec{r}|\ll l$,
about their mean positions 
on a triangular lattice.
For slow cracking on a frictional substrate, the motion of the grains 
is overdamped, the system evolves quasi-statically
with driving rate much slower than relaxation rate.  
In this case, we need not solve the equations of motion 
but instead may update the configurations 
according to threshold criteria specified below. 

In typical crack formations, the layer 
often hardens and/or weakens in time; it tends to contract
but is resisted by friction from the substrate. As a result,
stress is built up and relaxed slowly.
The simplest way to incorporate these effects is
first to pre-strain the array by
$s=(a-l)/a>0$, where $a\equiv 1$ and $l<a$ are the initial
and final grain spacing, respectively.
Then, two thresholds, $F_s$ for slipping and $F_c$ for cracking,
are decreased systematically to induce slipping and cracking.
This rule is more general than it seems:
As the evolution of the model 
is determined entirely by the ratios
between forces and thresholds, hardening, weakening
{\em and\/} their combination amount to this same rule.
Since the thresholds decrease by the same physical means, 
their ratio $\kappa\equiv F_c/F_s$ remains fixed.
In a drying experiment, the narrowing and breaking 
of liquid bridges between grains, due to evaporation, 
may provide a physical picture of this driving.
Even though there may not be an exact 
correspondence in experiments, and
other drivings are conceivable\cite{driving},
our choice is physically relevant and
sufficiently simple to allow for some analytic understanding.

The general expression of the force 
is rather complicated,
it results in slow updating.
Since in realistic situation the stress is primarily tensile 
and the strain $s\ll 1$, to a good approximation
we expand the force to first order in $\vec{r}$
to obtain a Hookean form for the force 
components 
on a block:
\begin{eqnarray}
F_x &=& \sum_{i=1}^6  
H_i [asC_i + (x_i-x)
(sS_i^2+C_i^2)+
\nonumber \\
&& \qquad
(y_i-y)(1-s)S_i C_i ],
\label{Eq:force}
\end{eqnarray}
where the index $i$ labels the six nearest neighbors 
and the associated bundles,
and the constants 
$C_i=\cos[(2i-1)\pi/6]$ and $S_i=\sin[(2i-1)\pi/6]$.
$F_y$ may be obtained from
$F_x$ by interchanges 
$x\leftrightarrow y$, and $C_i\leftrightarrow S_i$.

Initially, the system is globally stable.
In discrete simulation steps $t$,
it evolves according to these rules:
\begin{enumerate}
\item
The thresholds are lowered until 
either is exceeded somewhere in the system;
\item
if $F\equiv\sqrt{F_x^2+F_y^2}>F_s$, 
the block slips to a mechanically equilibrium position where $F\equiv 0$;
\item
if the bond tension $>F_c$ in a bundle of $H_i$ bonds,
$H_i\to H_i-1$ provided the bundle has not 
been damaged in the same step.
\end{enumerate}
The system relaxes until global stability is restored.
This constitutes one {\em simulation step or one event\/}.
In rule~3, we break no more than one bond/bundle/step 
in order to avoid instantaneous downward propagation of cracks
before the neighboring stress field has ever relaxed.
Otherwise, spurious long and vertical microcracks would be generated
which do not conform to real systems.
The progressive damage of the bundle mimics successive 
breakage of liquid bridges,
hence $H$ plays the role of the local thickness\cite{3D}.
Our algorithm is designed to take advantage of the 
dominance of the horizontal propagation of cracks,
due to the presence of the top, free surface.

\begin{figure}[htp]
\epsfig{figure=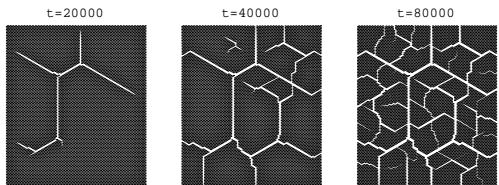,width=2.6in,angle=-0}
\caption{
Snapshots 
for $L=100$, $s=0.1$, $\kappa=0.5$, $H=9$.
Note the progression from straight to diffusive cracks.
}
\label{fig:conH9}
\end{figure}

{\em Results and Discussions} --
Throughout this work,
square systems are used with linear size $20 \leq L\leq 300$,
$s \leq 0.1$, and
free boundary conditions (FBC) are imposed\cite{BC}.
In the $\kappa-H$ subspace,
the time evolution can be divided into 
three qualitatively distinct phases:
\begin{enumerate}
\item[I.]
The system contracts by slippings, the slip size 
(total number of slippings in an event) grows in time; 
\item[II.]
There are slippings and bond breakings, 
the system is progressively damaged, then fragmented,
while contraction continues;
\item[III.]
Bond-breaking saturates, fragmentation stops, slipping
dominates as block spacing converges to the equilibrium value $l$.
\end{enumerate}
A typical evolution is shown in Fig.~\ref{fig:conH9}.
The main feature of phase I is the growth of correlations.
Since everything depends on $F_s$, 
we need to know how it varies first.
Its decay rate is governed by the density of 
$F$ near $F_s$ among the $ L^2$ blocks, where $0\leq F < F_s$.
To a first (mean-field) approximation, this density is $F_s/ L^2$.
Hence $F_s(t)-F_s(t+1)\propto F_s(t)/L^2$, 
leading to $F_s(t) \propto e^{-t/\tau}$
with a decay constant $\tau \sim L^z$ and 
a mean-field prediction $z_{\rm MF}=2$.
Numerically, we find $z=1.80 \pm 0.03$, 
reflecting the finite, albeit weak, spatial
correlation among the seeds that initiate distinct events.

\begin{figure}[htp]
\epsfig{figure=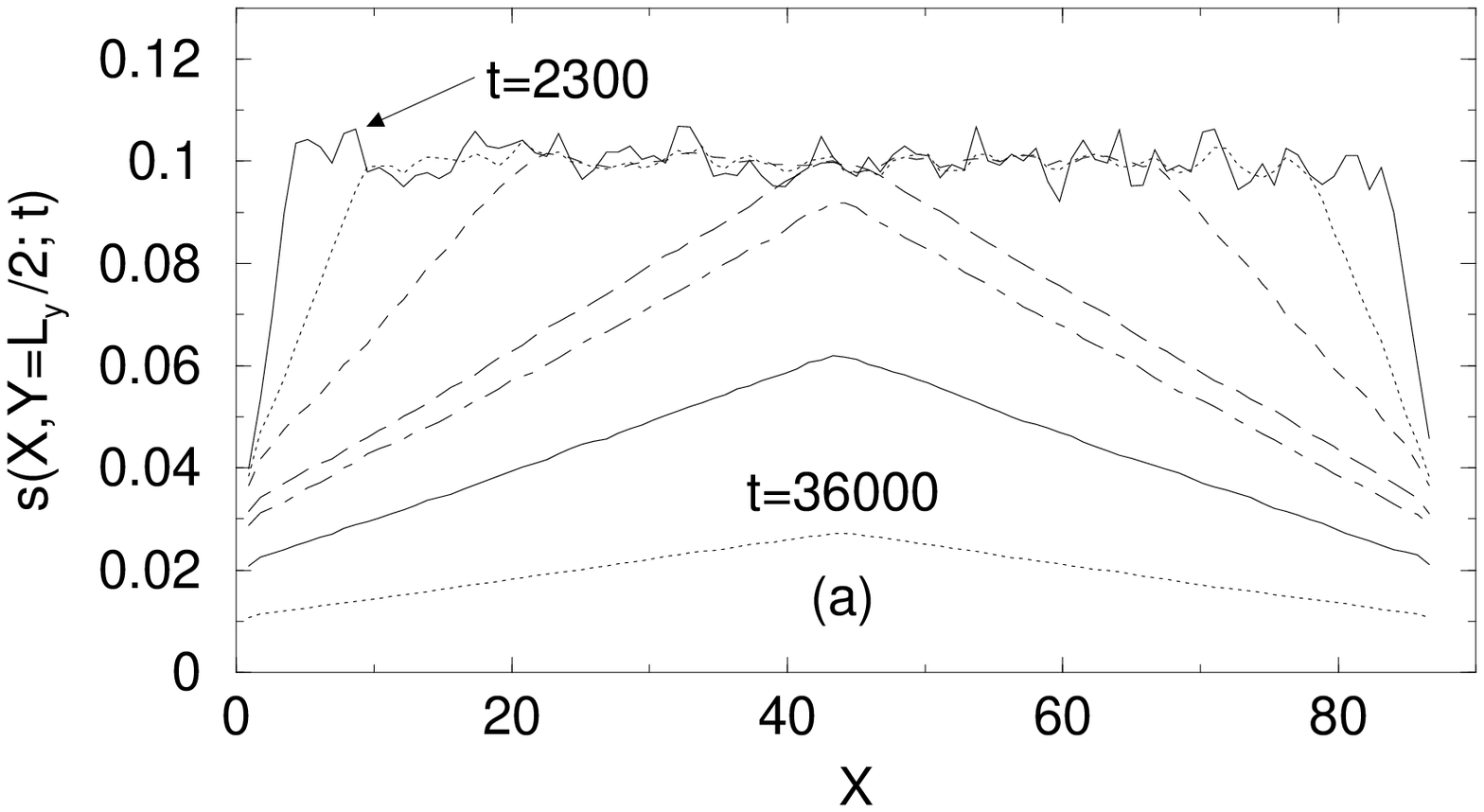,width=2.6in,angle=-00}
\epsfig{figure=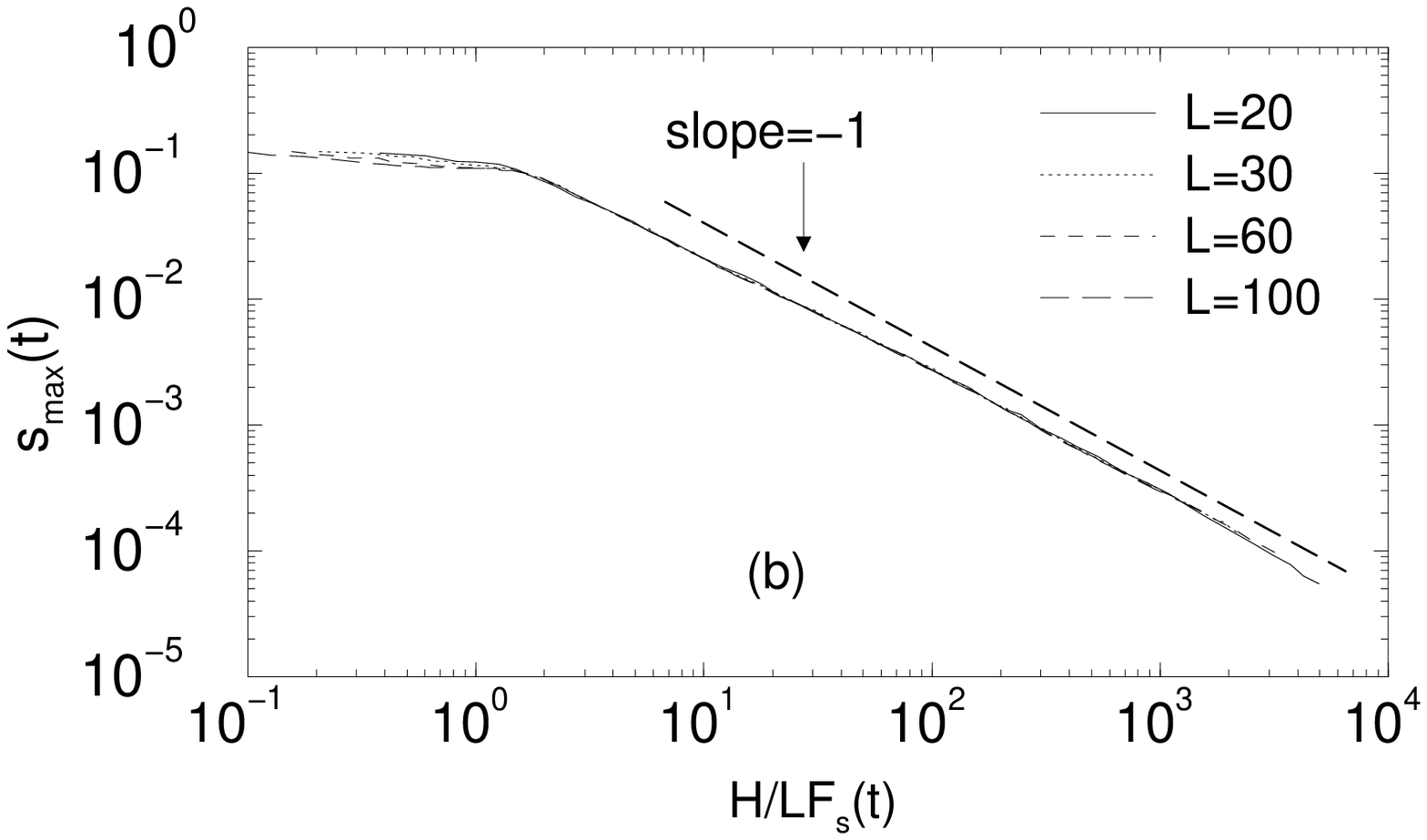,width=2.7in,angle=-00}
\caption{
(a) Strain field profiles show progressive strain relief as $t$ increases 
for $L=100$, $s=0.1$, $H=2$, $\kappa=\infty$.  It leads to
temporal variation of the maximum strain
in (b).
}
\label{fig:smax}
\end{figure}

Now we derive the growth laws for the strain field and the slip size.
Due to the diminishing threshold,
slippings start from the free edge and invade into the bulk,
giving rise to a strain-relieved peripheral region
of width denoted by $\xi$.
The strain increases from the 
boundary toward the bulk, reaching the bulk value $s$
(see Fig.~\ref{fig:smax}(a)).
To determine its profile, 
consider for the moment a one-dimensional version:
the force on the $i^{\rm th}$ block is given by
$F_i=(a-l+x_{i+1}-x_i)H-(a-l+x_i-x_{i-1})H=(s_i - s_{i-1}) H$. 
Hence $F=H ds/du$, where $u$ is the distance from the edge.
Metastability of the system requires
$ H ds/du < F_s$ but of the same order.
Thus we obtain for general dimensions the growth of the invasion length
\begin{equation}
\xi(t) \approx { \xi_0 s H \over 2 F_s(t)},
\label{Eq:xi_t}
\end{equation}
with constant $\xi_0 =O(1)$.
For a finite system of size $L$,  the invasion saturates
(i.e., $\xi=L$) when $F_s\leq F_s^{\rm sat}=\xi_0 s H/L$,
or equivalently $t\geq t_{\rm sat}=\tau \ln(H/F_s^{\rm sat})$.
As a result, the global maximum strain $s_{\rm max}$ 
in the system drops continuously thereafter
(see Fig.~\ref{fig:smax}(b)):
\begin{equation}
s_{\rm max}=\cases{s                     \quad& $t < t_{\rm sat}$\cr
                  {LF_s \over   \xi_0 H} \quad& $t > t_{\rm sat}.$\cr}
\\
\label{Eq:smax}
\end{equation}

\begin{figure}[htp]
\epsfig{figure=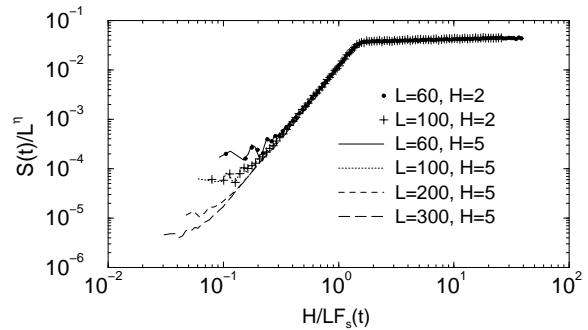,width=3.0in,angle=-00}
\caption{
Dynamic finite-size scaling 
of mean slip size toward criticality.
$s=0.1$, $\kappa=\infty$, $\eta=2.2\pm 0.03$.
}
\label{fig:Sfss}
\end{figure}

As $\xi(t)$ grows, so does the mean slip size, $S(t)$,
as required by energy balance.
The total energy relieved over the strain-relieved region
is of the order $U=\xi L s^2 H/2$. It is dissipated
by friction during slippings.  Since the slip distance is
$\alpha F_s/H$\cite{la97} 
where $\alpha=1/3(1+s)$\cite{redistribution},
each block slip dissipates an energy $\alpha F_s^2/2H$.
The total dissipation is then
$E=\int_0^t \,dt' S(t') \alpha F_s(t')^2/2H$.
Equating $E$ to $U$ and differentiating with respect to $t$,
we readily find 
\begin{equation}
S(t) = S_0 \left({H\over F_s(t)}\right)^3 \qquad t < t_{\rm sat},
\label{Eq:S_t}
\end{equation}
where the prefactor $S_0=\xi_0 s^3 L/ 2 \alpha \tau$.
Since the invasion is independent of $L$,
we have $\xi_0 \sim L^0$ and hence $S_0\sim L^{1-z}$.
For $t > t_{\rm sat}$, $S(t)$ saturates to a constant
$S_{\rm sat} = S_0 (L/s \xi_0)^3 \sim L^\eta$ which diverges 
as $L\to \infty$: the system becomes critical 
with a nontrivial exponent $\eta>0$.
Put together, we obtain a ``scaling relation" 
\begin{equation}
\eta=4-z, 
\label{Eq:eta_z}
\end{equation}
relating the dynamical and stationary behavior.
From $z_{\rm MF}=2$, we get $\eta_{\rm MF}=2$. 
Simulations give $\eta=2.20\pm 0.03$ instead, in 
agreement with the independent estimate of $z$.
From Eq.~(\ref{Eq:S_t}), it is evident that
dynamic scaling $S(t;H,L)=L^\eta \Phi(H/LF_s(t))$ is obeyed,
with the scaling function $\Phi(x\ll 1)\sim x^3$
and $\Phi(x\gg 1)=$const,
as exhibited in Fig.~\ref{fig:Sfss}.

Several remarks are in order:
(i) $\eta>2$ implies that each site slips an average 
$L^{\eta-2}$ times per event. Such repeated topplings
are manifestations of system-wide correlations in the force variable.
(ii) The connection between dynamics ($z$)
and criticality ($\eta$) is attributed to energy balance.
Thus we believe similar relations also exist for other forms of driving.
(iii) 
For finite $\kappa$, fracture eventually sets in 
before criticality is fully developed.  
However, even then, each fragment settles into a stationary, critical state 
characterized by the same $\eta$ and its fragment size.

Moving on to phase II, 
an important question in fracture is 
how its onset is affected by 
material properties and external conditions.
Apparently, fracture occurs when 
$\kappa F_s$ decreases below
the bond tension $s$ in the bulk, i.e., 
\begin{equation}
F_s \to F_s^* \approx {s\over \kappa}
\label{Eq:Fsc}
\end{equation}
from above.
Note that this result is independent of $H$, but the 
corresponding simulation step is: $t^* \approx \tau \ln(\kappa H/s)$,
because $F_s \approx He^{-t/\tau}$.
Hence, the substrate and the thickness 
have the same effect on the onset.
The extra waiting time 
for thicker layers or more frictional substrates
allows more slippings and hence stronger correlations at $t^*$ --
larger fragments are anticipated.


The difference in correlations is also reflected
in the morphology of cracks.
For very small $\kappa H$,
the substrate is very frictional or the layer very thin. 
Pinning is so strong that most avalanches are of size one.
In a virtually uncorrelated stress environment,
cracks soon appear but do not propagate.
Fragmentation is resulted from
percolation of microcracks.
For $\kappa H \ll 1$,  most bonds are ultimately broken to
turn the system into powder.

\begin{figure}[htp]
\epsfig{figure=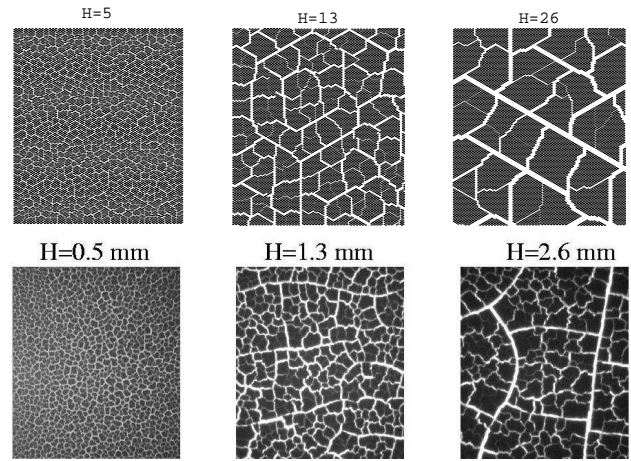,width=3.3in,angle=-00}
\caption{
Stationary patterns at different thickness.
Upper row: by simulation, $L=100$, $s=0.1$, $\kappa=0.25$. 
Increasing $\kappa$ at fixed $H$ has the same effect.
Lower row: from slow-drying experiment of cornstarch-water mixture
           with thickness matching that of simulation.
}
\label{fig:pattn}
\end{figure}

Going up in $\kappa H$, the larger friction or thickness
allows stronger correlations up to fracture,
measured by $\xi(t^*)$ and $S(t^*)$.
Relaxation of stress field around crack tips becomes more effective,
leading to more pronounced 
stress concentration and straight crack propagation.
The resulting fragmentation process is hierarchical:
after a cellular network is formed with the primary straight cracks,
secondary cracks break the fragments into smaller ones, and so on.
Cracks of later generations are more ``diffusive" in character, 
they tend to wiggle and branch under smaller stresses\cite{stress},
as there are more and more free boundaries.
These features are exemplified in Fig.~\ref{fig:pattn}.
At long times, fragmentation is complete, resulting in
fragments with areas that fit a log-normal distribution
(cf. \cite{la97}), and a mean area $A\sim H^2$ (see below).
Experimentally, morphology may be characterized by
the distribution $P(\theta)$ of crack angle $\theta$ at a joint.
A shift of the peak position from $\theta\approx 90^\circ$ to $60^\circ$ 
as thickness decreases
has been observed in amorphous media\cite{groisman}.
Although the usefulness of $P(\theta)$
is limited by lattice anisotropies in our model,
a direct comparison with real patterns (see Fig.~\ref{fig:pattn})
produced by slow drying of cornstarch-water mixtures
(to be reported elsewhere) reveals striking similarities.
Furthermore, the linear dependence of $A$ on thickness
agrees with previous experiments\cite{groisman,BBN}.

\begin{figure}[htp]
\epsfig{figure=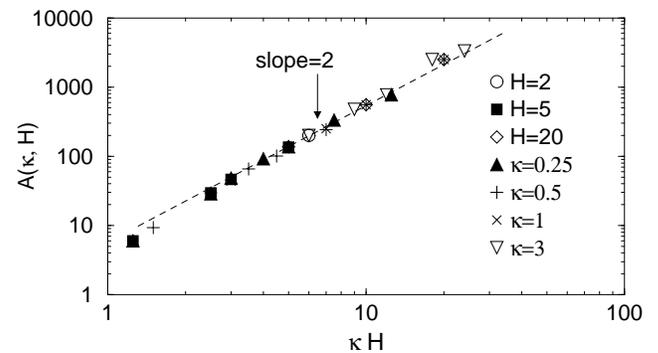,width=3.3in,angle=-00}
\caption{
Substrate-thickness scaling of mean fragment area. 
$L=100$, $s=0.1$, with either fixed $\kappa$ or $H$.
}
\label{fig:A_kH}
\end{figure}

As $\kappa H$ is increased further, the question of fracture or not arises.
Fracture is possible only if $\kappa F_s(t)<s_{\rm max}(t)$ at some $t$.
From Fig.~\ref{fig:smax}(b), this translates to 
the condition $\kappa F_s(t_{\rm sat}) \equiv s$ 
(equivalently $t_{\rm sat}\equiv t^*$).
A critical system size 
$L_c(\kappa,H)=\xi_0 \kappa H$ emerges,
below which a system never cracks at fixed $\kappa$ and $H$.
Equivalently, a finite system of size $L$ does not crack if
$\kappa H > L/\xi_0$.
This again agrees with the basic experimental fact that
a sufficiently small and thick layer does not crack.
Now, given that a system of size $L>L_c$ cracks, 
the next, even more interesting question of
pattern selection is when fragmentation stops.
We do not have a rigorous answer, as the dynamics
of interacting cracks is complicated,
but we argue that it stops when the
system has broken up into fragments of typical size $O(L_c)$.
This gives $A(\kappa,H)\sim (\kappa H)^2$.
This argument implies the memory 
of the correlations at the onset 
persists in the {\em final\/} fragmented state,
reminiscent of the generation of quenched disorder 
in systems with metastability and frustration\cite{memory} --
both features crucial to surface fracture.
Excellent scaling behavior in simulations (Fig.~\ref{fig:A_kH}) 
strongly supports this scenario.


In summary, we study a bundle-block model
for quasi-static, in-plane fracture.
Guided by experiments, 
we incorporate two factors that dictate the pattern-forming process --
the substrate and thickness -- 
by means of two control parameters.
The ensuing pattern, however, is characterized mostly by 
one scaling variable.
Self organization of the strain field prior to fracture is 
derived and shown to dictate the properties of the pattern, 
particularly to select the fragment size.
Our results answer, at least qualitatively, 
why thin layers crack finely and thick ones do not crack at all,
and why similarities are accompanied by a variability of scales.
Supported by experimental observations, 
our model appears to capture the salient features of real systems,
so that despite the use of a specific form of driving, 
the ideas may apply 
to other situations generally dominated by stick-slip motion.

We thank Yves Br\'echet and Eric To for useful discussions, 
and C.K. Chan for help on experiment.
This work is partially supported by the National Science Council of R.O.C.





\begin{references}

\bibitem[*]{email_ktl} E-mail: leungkt@phys.sinica.edu.tw

\bibitem[\dagger]{email_zn} E-mail: zneda@hera.ubbcluj.ro


\bibitem
{stevens} P.S. Stevens, ``Patterns in nature'', Atlantic Monthly Press (1974).

\bibitem
{lawn} B. Lawn, ``Fracture of brittle solids'', 2nd ed.,
Cambridge Univ. Press (1993).

\bibitem
{playa} 
J.T. Neal, A.M. Langer and P.F. Kerr,
Geol. Soc. Am. Bull., {\bf 79}, 69 (1968).

\bibitem{Walker}
J. Walker, Sci. Am. {\bf 255}, 178 (1986).

\bibitem{recent}
A. T. Skjeltorp and P. Meakin, Nature {\bf 335}, 424 (1988);
J. V. Andersen, Y. Br\'echet and H. J. Jensen,
Europhys. Lett. {\bf 26}, 13 (1994);
T. Hornig, I.M. Sokolov and A. Blumen, Phys. Rev. E {\bf 54}, 4293 (1996);
K.M. Crosby and R.M. Bradley, 
{\em ibid.} {\bf 55}, 6084 (1997);
J.V. Andersen, D. Sornette and K.-t. Leung, 
Phys. Rev. Lett. {\bf 78}, 2140 (1997).

\bibitem{groisman}
A. Groisman and E. Kaplan, Europhys. Lett. {\bf 25}, 415 (1994).

\bibitem{BBN}
Y. Br\'echet, D. Bellet and Z. N\'eda, Key Eng. Mat. {\bf 103}, 247 (1995).

\bibitem{la97}
K.-t. Leung and J. V. Andersen, 
Europhys. Lett. {\bf 38}, 589 (1997).

\bibitem{kitsunezaki}
S. Kitsunezaki, Phys. Rev. E {\bf 60}, 6449 (1999).


\bibitem{CZM}
J.S. Langer and A.E. Lobkovsky, 
J. Mech. Phys. Solids {\bf 46}, 1521 (1998); and refs. therein. 

\bibitem{Kachanov}
M. Kachanov, 
Adv. Appl. Mech. {\bf 30}, 259-438 (1994).

\bibitem{HR} 
H.J. Herrmann and S. Roux, eds., ``Statistical models for the fracture
of disordered media'', North-Holland (1990).

\bibitem{driving} 
If the thresholds vary independently, 
fine tuning becomes necessary to produce
nontrivial steady-state patterns\cite{la97}; 
no such tuning is evident in nature.

\bibitem{Hutchinson} 
J.W. Hutchinson, 
Adv. in Appl. Mech. {\bf 29}, 63 (1992).

\bibitem{3D}
This is a trick to introduce the {\em effect\/} 
of thickness without incurring
a demanding three-dimensional simulation.

\bibitem{BC} 
Partially and fully periodic boundary conditions are also
considered. They lead to similar patterns with the same
fragment-area statistics.

\bibitem{redistribution}
$\alpha$ specifies the fraction of the force components 
distributed to neighbors in a slip. 
Note that without external driving, 
all force components are strictly conserved locally
during slipping {\em and\/} cracking, for all $s$.

\bibitem{stress}
Smaller initial strain leads to more diffusive cracks. 
Generally, typical straight-crack length increases with $s$.


\bibitem{memory}
S. Boettcher and M. Paczuski, Phys. Rev. Lett. {\bf 79}, 889 (1997);
and refs. therein.


\end{references}
\end{document}